\documentstyle[12pt]{article}
\newcommand{\bfp}{\mbox{\boldmath $p$}}

\newcommand{\bfk}{\mbox{\boldmath $k$}}
\newcommand{\bfh}{\mbox{\boldmath $h$}}
\newcommand{\bfq}{\mbox{\boldmath $q$}}
\newcommand{\bfy}{\mbox{\boldmath $y$}}

\def\nostrocostruttino#1\over#2{\mathrel{\mathop{\kern 0pt \rlap
{\hbox{$#1$}}} \hbox{\kern-.135em $#2$}}}

\newcommand{\NP}[1]{{\it Nucl.\ Phys.}\ {\bf #1}}
\newcommand{\ZP}[1]{{\it Z.\ Phys.}\ {\bf #1}}
\newcommand{\PL}[1]{{\it Phys.\ Lett.}\ {\bf #1}}
\newcommand{\PR}[1]{{\it Phys.\ Rev.}\ {\bf #1}}

\newcommand{\beq}{\begin{equation}}
\newcommand{\eeq}{\end{equation}}
\newcommand{\barr}{\begin{eqnarray}}
\newcommand{\earr}{\end{eqnarray}}
\newcommand{\ba}{\begin{array}}
\newcommand{\ea}{\end{array}}
\newcommand{\ee}{e^-e^+}
\newcommand{\qq}{q\bar q}
\newcommand{\uu}{u\bar u}
\newcommand{\dd}{d\bar d}
\newcommand{\la}{\lambda}

\begin{document}

\begin{center}
{\LARGE\bf Single spin asymmetries in
{\mbox{\boldmath $\ell p^\uparrow \!\! \to hX$}} \\
and {\mbox{\boldmath $e^-e^+ \to q\bar q \to h^\uparrow X$}}} \\

\vspace{1cm}
{M. Anselmino}

\vspace*{0.5cm}

{\it Dipartimento di Fisica Teorica, Universit\`a di Torino and \\
      INFN, Sezione di Torino, Via P. Giuria 1, 10125 Torino, Italy} \\

\vspace*{1cm}
\end{center}

\begin{abstract}
Spin observables may reveal much deeper properties of non perturbative 
hadronic physics than unpolarized quantities. We discuss here possible 
origins of single spin asymmetries in DIS, absent in the elementary 
lepton-quark interactions, and suggest strategies to isolate and understand 
some non perturbative spin dependence of distribution and fragmentation 
functions. We also discuss the 
polarization of hadrons produced in $e^+e^-$ annihilation at LEP and 
show how final state $\qq$ interactions may give origin 
to non zero values of the off-diagonal element $\rho^{\,}_{1,-1}$ 
of the helicity density matrix of vector mesons: some predictions are given 
for $K^{*0}, \phi$ and $D^*$ in agreement with recent OPAL data. 
Possible analogous effects in DIS and other processes are suggested. 
\end{abstract}

\section{Introduction}

\vspace{1mm}
\noindent

The spin properties of hadrons inclusively produced in high energy 
interactions are related to the fundamental properties of quarks and 
gluons and to their elementary interactions in a much more subtle way 
than unpolarized quantities. They test unusual basic dynamical 
properties and reveal how the usual  models for quark distribution and
hadronization -- successful in predicting unpolarized cross-sections -- may 
not be adequate to describe spin effects. We consider here such two cases,
single spin asymmetries in Deep Inelastic Scattering and the polarization
of mesons produced in the fragmentation of polarized quarks at LEP.

\section{Single spin asymmetries in DIS}

\vspace{1mm}
\noindent

We discuss first single spin asymmetries in DIS processes. 
We start by reminding that single spin asymmetries in large $p_{_T}$ inclusive 
hadronic reactions are forbidden in leading-twist perturbative QCD, 
reflecting the fact that single spin asymmetries are zero at the partonic 
level and that collinear parton configurations inside hadrons do not allow 
single spin dependences. However, experiments tell us in several cases, 
\cite{ada1,ada2} that single spin asymmetries are large and indeed non 
negligible.

The usual arguments to explain this apparent disagreement between pQCD 
and experiment invoke the moderate $p_{_T}$ values of the data -- a few 
GeV, not quite yet in the true perturbative regime -- and the importance 
of higher-twist effects. Several phenomenological models have recently
attempted to explain the large single spin asymmetries observed in
$p^\uparrow p \to \pi X$ as twist-3 effects which might be due to intrinsic 
partonic $\bfk_\perp$ in the fragmentation \cite{col1} and/or distribution 
functions \cite{siv1}-\cite{noi}. 

Let us consider a process in which one has convincing evidence 
that partons and perturbative QCD work well and successfully describe the 
unpolarized and leading-twist spin data, namely Deep Inelastic Scattering 
(DIS). In particular we shall discuss single spin asymmetries in the inclusive,
$\ell N^\uparrow \to \ell + jets$ and $\ell N^\uparrow \to hX$, 
reactions looking at possible origins of such asymmetries and 
devising strategies to isolate and discriminate among them \cite{alm}.

According to the QCD hard scattering picture and the factorization theorem 
\cite{col1,col2,col3} the cross-section for the $\ell N^\uparrow \to hX$ 
reaction is given by
\barr
& & \frac{E_h \, d^3\sigma^{\ell + N,S \to h + X}} {d^{3} \bfp_h} = 
\sum_{q; \la^{\,}_{q^{\prime}}, \la^{\prime}_{q^{\prime}}, \la^{\,}_h} 
\int \frac {dx \, d^2\bfk_\perp d^2\bfk_\perp^\prime} {\pi z} \label{gen} \\  
& & \tilde f_{q/N}^{N,S}(x, \bfk_\perp) \>
\frac{d\hat\sigma^{q,P_q}}{d\hat t}(x, \bfk_\perp, \bfk_\perp^\prime) \>
\rho^{q^{\prime}}_{\la^{\,}_{q^{\prime}}, \la^{\prime}_{q^{\prime}}}
(x, \bfk_\perp, \bfk_\perp^\prime) \> \widetilde 
D_{\la^{\,}_h, \la^{\,}_h}^{\la^{\,}_{q^{\prime}}, \la^{\prime}_{q^{\prime}}}
(z, \bfk_\perp^\prime) \,. \nonumber
\earr

Let us briefly discuss the different quantities appearing in the above 
equation; more details can be found in Ref. \cite{alm}.
$\tilde f_{q/N}^{N,S}(x, \bfk_\perp)$ is the quark distribution function,
that is the total number density of quarks $q$  with momentum fraction 
$x$ and intrinsic transverse momentum $\bfk_\perp$ inside a polarized nucleon
$N$ with spin four-vector $S$. 

$d\hat\sigma^{q,P_q} / d\hat t$ is the cross-section for the $\ell 
q^\uparrow \to \ell q$ process, with an unpolarized lepton and an initial 
quark with polarization $P_q$, while the final quark and lepton polarization
are summed over. Notice that for helicity conserving elementary 
interactions $d\hat\sigma^{q,P_q} / d\hat t$ equals the unpolarized 
cross-section $d\hat\sigma^q / d\hat t$.
$\rho^{q^\prime}$ is the helicity density matrix of the final quark 
produced in the $\ell q^\uparrow$ interaction and 
\beq
\widetilde D^h_{q, s_q}(z, \bfk_\perp) = \sum_{\la^{\,}_q, \la^{\prime}_q} 
\rho^q_{\la^{\,}_q \la^{\prime}_q} \> \widetilde D_{\la^{\,}_h, 
\la^{\,}_h}^{\la^{\,}_q, \la^{\prime}_q}(z, \bfk_\perp)
\eeq
describes the fragmentation process of a polarized quark $q$ 
with spin $s_q$ into a hadron $h$ with helicity $\la^{\,}_h$, 
momentum fraction $z$ and intrinsic transverse momentum $\bfk_\perp$ with 
respect to the jet axis. It is simply the inclusive cross-section for
the $q^\uparrow \to hX$ fragmentation process. As it will be shown in the 
second part we can safely neglect the coherent interactions of the 
fragmenting quark, as we are not looking at the spin state of the final 
hadron. The usual unpolarized fragmentation function is given by
\beq
D^h_q(z) = {1\over 2} \sum_{\la^{\,}_q, \la^{\,}_h}
\int d^2\bfk_\perp \>
\widetilde D^{\la^{\,}_q,\la^{\,}_q}_{\la^{\,}_h,\la^{\,}_h}
(z, \bfk_\perp) \,.
\label{fr}
\eeq

Similar formulae hold also when the elementary interaction
is $\ell q \to \ell q g$ rather than $\ell q \to \ell q$: in the latter 
case two jets are observed in the final state -- the target jet and the 
current quark jet -- and in the former case three -- the target jet and 
$q$ + $g$ current jets.

In Eq. (\ref{gen}) we have taken into account intrinsic transverse momenta 
both in the distribution and the fragmentation process;
the $\bfk_\perp$ dependences are expected to have negligible effects on 
unpolarized variables for which they are indeed usually neglected, but 
they can be of crucial importance for some single spin observables, as 
discussed in Refs. \cite{col1}-\cite{noi}.

We discuss now possible sources of single spin effects in Eq. (\ref{gen}).

\goodbreak
\vskip 12pt
\noindent
{\mbox{\boldmath $k_\perp$}} {\bf effects in fragmentation process}
\cite{col1}
\vskip 6pt
\nobreak
 
The fragmentation process of a transversely polarized quark into a hadron 
$h$ (whose polarization in not observed) with fixed $z$ and $\bfk_\perp$ 
may depend on the quark spin orientation, provided the quark spin 
$s_q$ has a component perpendicular to the hadron-quark plane (otherwise 
any spin dependence would be forbidden by parity conservation). That is, 
there might be a non zero {\it quark analysing power} \cite{col1}:
\beq
A_q^h \equiv {\widetilde D^h_{q, s_q}(z, \bfk_\perp) 
- \widetilde D^h_{q,-s_q}(z, \bfk_\perp) \over \widetilde 
D^h_{q, s_q}(z, \bfk_\perp) + \widetilde D^h_{q,-s_q}(z,\bfk_\perp)} \,\cdot
\label{qap}
\eeq
By rotational invariance $\widetilde D^h_{q,-s_q}(z,\bfk_\perp) =
\widetilde D^h_{q, s_q}(z,-\bfk_\perp)$, which shows immediately how the 
quark analysing power vanishes for $\bfk_\perp = 0$. 
 
\goodbreak
\vskip 12pt
\noindent
{\mbox{\boldmath $k_\perp$}} {\bf effects in distribution functions}
\cite{siv1}-\cite{noi}
\vskip 6pt
\nobreak

A similar idea had been previously proposed in Refs. \cite{siv1, siv2}
and later rediscovered in Ref. \cite{noi}, concerning the distribution
functions: that is, the number of quarks (whose spin is not observed)
with fixed $x$ and $\bfk_\perp$ inside a transversely polarized nucleon
may depend on the nucleon spin orientation and the function
\beq
\Delta\tilde f_{q/N}^{N^\uparrow}(x, \bfk_\perp) \equiv
\tilde f_{q/N}^{N^\uparrow}(x, \bfk_\perp) - 
\tilde f_{q/N}^{N^\downarrow}(x, \bfk_\perp)
= \tilde f_{q/N}^{N^\uparrow}(x, \bfk_\perp) - 
\tilde f_{q/N}^{N,^\uparrow}(x,-\bfk_\perp)
\label{nap}
\eeq
may be different from zero. $\uparrow$ and $\downarrow$ refer to the 
nucleon spin up or down with respect to the quark-nucleon plane. 

In terms of the usual light-cone operator definition of structure 
functions one has \cite{col1,dra}
\barr
\Delta\tilde f_{q/N}^{N^\uparrow}
&=& 2 \, {\rm Im} \int {dy^- d\bfy_\perp \over(2 \pi)^3}
e^{-i x p^+ y^- + i \bfk_\perp\cdot\bfy_\perp} \nonumber\\
&&\quad\langle p, -|\bar\psi_a(0,y^-,y_\perp)
{\gamma^+\over 2}\psi_a(0)|p, + \rangle \,. 
\label{del+-}
\earr

In Ref. \cite{col1} it is argued that such off-diagonal (in the helicity
basis) matrix elements are zero due to the time-reversal invariance of QCD, 
and indeed this is proven by exploiting the time-reversal and parity 
transformation properties of free Dirac spinors. However, in Ref. \cite{dra} 
it has been shown that this need not be so in chiral models with quark moving 
in a background of chiral fields.

Both $\widetilde D^h_{q^\uparrow}(z, \bfk_\perp) 
- \widetilde D^h_{q^\downarrow}(z, \bfk_\perp)$ and 
$\tilde f_{q/N}^{N^\uparrow}(x, \bfk_\perp) - 
\tilde f_{q/N}^{N^\downarrow}(x, \bfk_\perp)$ can be considered as 
new fundamental spin and $\bfk_\perp$ dependent non perturbative 
functions describing respectively quark fragmentation and distribution
properties. In the sequel we shall devise strategies to test their
relevance. 

\vskip 12pt
\noindent
{\bf Single spin effects in the elementary interactions} 
\vskip 6pt

As we already mentioned both perturbative QED and QCD at high energy 
do not allow single helicity flips in the $\ell q$ interactions, so that 
there cannot be any dependence on the quark polarization in 
$d\hat\sigma^{q,P_q}/d\hat t$. Similarly, the perturbative QCD evolution
of the distribution and fragmentation functions is not expected to
introduce any single spin dependence. We must conclude that the hard 
elementary interactions are unlikely to introduce any single
spin effect: however, this basic QED and QCD property should also
be tested. 

\vskip 6pt
Let us now describe a set of possible measurements which could single out
some of the above mechanisms and test them. 

\goodbreak
\vskip 6pt
\noindent
$a) \> \ell N^\uparrow \to \ell + 2\,jets$
\vskip 4pt
\nobreak

Here one avoids any fragmentation effect by looking at the fully 
inclusive cross-section for the process $\ell N^\uparrow \to \ell + 2\, jets$,
the 2 jets being the target and current ones; this is the usual DIS, 
and Eq. (\ref{gen}) becomes  
\beq
\frac{d^2\sigma^{\ell + N,S \to \ell + X}} {dx \, dQ^2} = \sum_q
\int d^2\bfk_\perp \> \tilde f_{q/N}^{N,S}(x, \bfk_\perp) \>
\frac{d\hat\sigma^{q,P_q}}{d\hat t}(x, \bfk_\perp) \,. 
\label{gena}
\eeq

In this case the elementary interaction is a pure QED, 
helicity conserving one, $\ell q \to \ell q$, and $d\hat\sigma^{q,P_q}/d\hat t$
cannot depend on the quark polarization. Some spin dependence might only
remain in the distribution function, due to intrinsic $\bfk_\perp$ effects
\cite{siv1}-\cite{noi}, \cite{dra} and we have
\barr
\frac{d^2\sigma^{\ell N^\uparrow \to \ell + X}} {dx \, dQ^2} 
&-& \frac{d^2\sigma^{\ell N^\downarrow \to \ell + X}} {dx \, dQ^2} 
= \sum_q \int d^2\bfk_\perp \nonumber \\
&\times& \Delta\tilde f_{q/N}^{N^\uparrow}(x, \bfk_\perp) \>\>
\frac{d\hat\sigma^q}{d\hat t}(x, \bfk_\perp) \,. 
\label{asymg}
\earr
Despite the fact that $\Delta\tilde f_{q/N}^{N^\uparrow}$ is an odd
function of $\bfk_\perp$ a non zero value of the above difference 
-- of ${\cal O}(k_\perp/\sqrt {Q^2})$, twist 3 -- 
might remain even after integration on $d^2\bfk_\perp$ because of
the $\bfk_\perp$ dependence of $d\hat\sigma^q/d\hat t$, similarly 
to what happens in $pp^\uparrow \to \pi X$ \cite{noi}. The observation
of a non vanishing value of the single spin effect of Eq. (\ref{asymg})
would be a decisive test in favour of the mechanism suggested in 
Refs. \cite{siv1}-\cite{noi} and would allow an estimate of the
new function (\ref{nap}).

\vskip 6pt
\noindent
$b) \> \ell N^\uparrow \to h + X \> (2\,jets, \> \bfk_\perp \not= 0)$
\vskip 4pt

One looks for a hadron $h$, with transverse momentum $\bfk_\perp$, inside 
the quark current jet; the final lepton may or may not be observed.
The elementary subprocess is $\ell q \to \ell q$ and Eq. (\ref{gen}) yields
\barr
& & \frac{E_h \, d^5\sigma^{\ell + N^\uparrow \to h + X}} 
{d^{3} \bfp_h d^2 \bfk_\perp} 
- \frac{E_h \, d^5\sigma^{\ell + N^\downarrow \to h + X}} 
{d^{3} \bfp_h d^2 \bfk_\perp} \label{coll} \\
&=& \sum_q \int \frac {dx} {\pi z} \>    
\Delta_{_T} q(x) \> \Delta_{_N} \hat\sigma^q (x, \bfk_\perp) \,
\left[ \tilde D^h_{q^\uparrow}(z, \bfk_\perp)
- \tilde D^h_{q^\uparrow}(z, - \bfk_\perp) \right]
\nonumber
\earr
where $\Delta_{_T}q$ is the polarized number density for transversely spinning 
quarks $q$ and $\Delta_{_N} \hat\sigma^q$ is the elementary cross-section 
double spin asymmetry
\beq
\Delta_{_N} \hat\sigma^q = {d\hat \sigma^{\ell q^\uparrow \to 
\ell q^\uparrow} \over d\hat t} - {d\hat \sigma^{\ell q^\uparrow \to 
\ell q^\downarrow} \over d\hat t} \,\cdot
\label{del}
\eeq

In Eq. (\ref{coll}) we have neglected the $\bfk_\perp$ effect in the 
distribution function, which can be done once the asymmetry discussed
in $a)$ turns out to be negligible. We are then testing directly the 
mechanism suggested in Ref. \cite{col1} and a non zero value of
the l.h.s. of Eq. (\ref{coll}) would be a decisive test in its favour 
and would allow an estimate of the new function appearing in the
numerator of Eq. (\ref{qap}). Notice again that even upon integration over
$d^2\bfk_\perp$ the spin asymmetry of Eq. (\ref{coll}) might survive,
at higher twist order $k_\perp/p_{_T}$, due to some $\bfk_\perp$ dependence 
in $\Delta_{_N} \hat\sigma^q$.

\goodbreak
\vskip 6pt
\noindent
$c) \> \ell N^\uparrow \to h + X \> (2\,jets, \> \bfk_\perp = 0)$
\vskip 4pt
\nobreak

By selecting events with the final hadron collinear to the jet axis
($\bfk_\perp = 0$) one forbids any single spin effect in the fragmentation
process. As in the fully inclusive case $a)$ the observation of a
single spin asymmetry in such a case would imply single, $\bfk_\perp$
dependent, spin effects in the distribution functions. 

\vskip 6pt
\noindent
$d) \> \ell N^\uparrow \to h + X \> (3\,jets, \> \bfk_\perp \not= 0)$
\vskip 4pt
 
The elementary process is now $\ell q \to \ell qg$ and one looks at 
hadrons with $\bfk_\perp \not= 0$ inside the $q$ current jet. Single
spin asymmetries can originate either from single spin effects in the 
fragmentation process or distribution functions. One should not, in 
principle, forget possible spin effects in the elementary QCD interaction.

\vskip 6pt
\noindent
$e) \> \ell N^\uparrow \to \ell + 3\,jets$ or 
$\ell N^\uparrow \to h + X \> (3\,jets, \> \bfk_\perp = 0)$
\vskip 4pt

These cases are analogous to $a)$ and $c)$ respectively: the measurement 
eliminates spin effects arising from the fragmentation functions. 
The only possible origin of a single spin asymmetry would reside 
in the distribution function. However, if no effect is observed in cases
$a)$ and $c)$, but some effect is observed here, then one has to 
conclude that there must be some single spin effect in the elementary 
QCD interaction. Utterly unexpectedly, this would question the
validity of quark helicity conservation, a fundamental property of
pQCD which has never been directly tested. 

\vskip 6pt
In summary, a study of single transverse spin asymmetries in DIS could
provide a series of profound tests of our understanding of large $p_{_T}$
QCD-controlled reactions.

\section{{\mbox{\boldmath $\rho^{\,}_{1,-1}(V)$}} in the process 
{\mbox{\boldmath $e^- e^+ \to q\bar q \to V + X$}}}

\vspace{1mm}
\noindent

We consider now the spin properties of hadrons produced at LEP. It was pointed 
out in Refs. \cite{akp} and \cite{aamr} that final state interactions 
between the $q$ and $\bar q$ created in $e^+ e^-$ annihilations 
-- usually neglected, but indeed necessary -- might give origin to non 
zero spin observables which would otherwise be forced to vanish: 
the off-diagonal element $\rho^{\,}_{1,-1}$ of the helicity density matrix of 
vector mesons may be sizeably different from zero \cite{akp} due to a 
coherent fragmentation process which takes into account $q \bar q$ 
interactions. The incoherent fragmentation of a single independent quark 
leads instead to zero values for such off-diagonal elements. 

We present predictions \cite{abmq} for $\rho^{\,}_{1,-1}$ of several vector 
mesons $V$ provided they are produced in two jet events, carry
a large momentum or energy fraction $z=2E_{_V}/\sqrt s$, and have a small
transverse momentum $p_{_T}$ inside the jet. Our estimates are in agreement
with the existing data and are crucially related both to the presence 
of final state interactions and to the Standard Model couplings of the 
elementary $e^- e^+ \to q \bar q$ interaction. 

The helicity density matrix of a hadron $h$ inclusively produced in the 
two jet event $e^- e^+ \to q\bar q \to h + X$ can be written as 
\cite{akp, aamr}
\beq
\rho^{\,}_{\la^{\,}_h \la^{\prime}_h}(h) 
= {1\over N_h} \sum_{q,X,\la^{\,}_X,\la^{\,}_q,\la^{\,}_{\bar q},
\la^{\prime}_q,\la^{\prime}_{\bar q}} 
D^{\,}_{\la^{\,}_h \la^{\,}_X; \la^{\,}_q,\la^{\,}_{\bar q}} \>\>
\rho^{\,}_{\la^{\,}_q,\la^{\,}_{\bar q};
\la^{\prime}_q,\la^{\prime}_{\bar q}}\,(\qq) \>\> 
D^*_{\la^{\prime}_h \la^{\,}_X; \la^{\prime}_q,\la^{\prime}_{\bar q}} \,,
\label{rhoh}
\eeq
where $\rho(\qq)$ is the helicity density matrix of the $q\bar q$ state 
created in the annihilation of the unpolarized $e^+$ and $e^-$,
\beq
\rho^{\,}_{\la^{\,}_q,\la^{\,}_{\bar q};
\la^{\prime}_q,\la^{\prime}_{\bar q}}\,(\qq)
= {1\over 4N_{\qq}} \sum_{\la^{\,}_{-}, \la^{\,}_{+}}
M^{\,}_{\la^{\,}_q \la^{\,}_{\bar q};\la^{\,}_{-} \la^{\,}_{+}} \>
M^*_{\la^{\prime}_q \la^{\prime}_{\bar q}; \la^{\,}_{-} \la^{\,}_{+}} \,.
\label{rhoqq}
\eeq
The $M$'s are the helicity amplitudes for the $\ee \to \qq$ process and
the $D$'s are the fragmentation amplitudes, {\it i.e.} the helicity
amplitudes for the process $\qq \to h+X$; the $\sum_{X,\lambda_X}$ stands 
for the phase space integration and the sum over spins of all the unobserved 
particles, grouped into a state $X$. The normalization factors $N_h$ and
$N_{\qq}$ are given by:
\beq
N_h = \sum_{q,X; \la^{\,}_h, \la^{\,}_X, \la^{\,}_q, \la^{\,}_{\bar q},
\la^{\prime}_q, \la^{\prime}_{\bar q}} 
D^{\,}_{\la^{\,}_h \la^{\,}_X; \la^{\,}_q,\la^{\,}_{\bar q}} \>\>
\rho^{\,}_{\la^{\,}_q,\la^{\,}_{\bar q};
\la^{\prime}_q,\la^{\prime}_{\bar q}}\,(\qq) \>\> 
D^*_{\la^{\,}_h \la^{\,}_X; \la^{\prime}_q,\la^{\prime}_{\bar q}} \,
= \sum_q D^h_q \,,
\label{nh}
\eeq
where $D^h_q$ is the usual fragmentation function of quark $q$ into 
hadron $h$, and 
\beq
N_{\qq} = {1\over 4} 
\sum_{\la^{\,}_q, \la^{\,}_{\bar q}; \la^{\,}_{-}, \la^{\,}_{+}} \vert 
M^{\,}_{\la^{\,}_q \la^{\,}_{\bar q}; \la^{\,}_{-} \la^{\,}_{+}} \vert^2 \,.
\label{nqq}
\eeq

The helicity density matrix for the $q\bar q$ state can be computed 
in the Standard Model and its non zero elements are given by
\barr
\rho^{\,}_{+-;+-}(\qq) &=& 1 - \rho^{\,}_{-+;-+}(\qq) \> \simeq \> 
{1\over 2}\,{(g_{_V} - g_{_A})^2_q \over (g^2_{_V} + g^2_{_A})_q}
\label{rhoqqd} \\
\rho^{\,}_{+-;-+}(\qq) &=& \rho^*_{+-;-+}(\qq) \> \simeq \> 
{1\over 2}\,{(g^2_{_V} - g^2_{_A})_q \over (g^2_{_V} + g^2_{_A})_q} 
\, {\sin^2\theta \over 1+ \cos^2\theta} \, \cdot
\label{rhoqqap}
\earr
These expressions are simple but approximate and hold at the $Z_0$ pole, 
neglecting electromagnetic contributions, masses and terms proportional 
to $g_{_V}^l$; the full correct expressions can be found in Ref. \cite{abmq}.

Notice that, inserting the values of the coupling constants
\beq
g_{_V}^{u,c,t} =  \>\> {1\over 2} - {4\over 3}\sin^2\theta_{_W} \quad\quad
g_{_V}^{d,s,b} = -{1\over 2} + {2\over 3}\sin^2\theta_{_W} \quad\quad
g_{_A}^{u,c,t} = - g_{_A}^{d,s,b} = {1\over 2} \label{cc}
\nonumber
\eeq 
one has
\beq
\rho^{\,}_{+-;-+}(\uu, c\bar c, t\bar t) 
\simeq -0.36 {\sin^2\theta \over 1 + \cos^2\theta} 
\quad\quad 
\rho^{\,}_{+-;-+}(\dd, s\bar s, b\bar b) 
\simeq -0.17 {\sin^2\theta \over 1 + \cos^2\theta} 
\, \cdot \label{rho+-ap}
\eeq

Eq. (\ref{rho+-ap}) clearly shows the $\theta$ dependence of 
$\rho^{\,}_{+-;-+}(\qq)$. In case of pure electromagnetic interactions
($\sqrt s \ll M_{_Z}$) one has exactly:
\beq
\rho^{\gamma}_{+-;-+}(\qq) = {1\over 2}
\,{\sin^2\theta \over 1+ \cos^2\theta} \, \cdot
\label{rhoqqelm}
\eeq
Notice that Eqs. (\ref{rho+-ap}) and (\ref{rhoqqelm}) have the same
angular dependence, but a different sign for the coefficient in front,
which is negative for the $Z$ contribution.

By using the above equations for $\rho(\qq)$ into Eq. (\ref{rhoh}) one 
obtains the most general expression of $\rho(h)$ in terms of the $\qq$ spin 
state and the unknown fragmentation amplitudes.

Despite the ignorance of the fragmentation process some predictions
can be made \cite{abmq} by considering the production of hadrons almost 
collinear with the parent jet: the $q \bar q \to h + X$ fragmentation is 
then essentially a c.m. forward process and the unknown $D$ amplitudes must 
satisfy the angular momentum conservation relation \cite{bls}
\beq
D^{\,}_{\la^{\,}_h \la^{\,}_X; \la^{\,}_q,\la^{\,}_{\bar q}} 
\sim \left( \sin{\theta_h\over 2} \right)^{
\vert \la^{\,}_h - \la^{\,}_X - \la^{\,}_q + \la^{\,}_{\bar q} \vert} 
\simeq \left( {p_{_T}\over z\sqrt s} \right)^{
\vert \la^{\,}_h - \la^{\,}_X - \la^{\,}_q + \la^{\,}_{\bar q} \vert} \,,
\label{frd}
\eeq
with $\theta_h$ the angle between the hadron momentum, 
$\bfh = z \bfq + \bfp_{_T}$, and the quark momentum $\bfq$.

The bilinear combinations of fragmentation amplitudes contributing to
$\rho(h)$ are then not suppressed by powers of $(p_{_T}/z\sqrt s)$
only if the exponent in Eq. (\ref{frd}) is zero, which greatly reduces the 
number of relevant helicity configurations.

The fragmentation process is a parity conserving one and the fragmentation
amplitudes must then also satisfy the forward parity relationship
\beq
D^{\,}_{-\la^{\,}_h -\la^{\,}_X; -+} = (-1)^{S^{\,}_h + S^{\,}_X + 
\la^{\,}_h - \la^{\,}_X} \> 
D^{\,}_{\la^{\,}_h \la^{\,}_X; +-} \,.
\label{par}
\eeq

Before presenting analytical and numerical results for the coherent quark 
fragmentation let us remember that in case of incoherent single quark 
fragmentation Eq. (\ref{rhoh}) becomes
\beq
\rho^{\,}_{\la^{\,}_h \la^{\prime}_h}(h) 
= {1\over N_h} \sum_{q,X,\la^{\,}_X,\la^{\,}_q,\la^{\prime}_q}
D^{\,}_{\la^{\,}_h \la^{\,}_X; \la^{\,}_q} \>\>
\rho^{\,}_{\la^{\,}_q \la^{\prime}_q} \>\>
D^*_{\la^{\,}_h \la^{\,}_X; \la^{\,}_q} \,,
\label{rhohp1}
\eeq
where $\rho(q)$ is the quark $q$ helicity density matrix related to
$\rho(\qq)$ by
\beq
\rho^{\,}_{\la^{\,}_q \la^{\prime}_q} = \sum_{\la^{\,}_{\bar q}}
\rho^{\,}_{\la^{\,}_q, \la^{\,}_{\bar q}; \la^{\prime}_q,
\la^{\,}_{\bar q}} (\qq) \,.
\eeq

In such a case angular momentum conservation for the collinear quark 
fragmentation requires $\la^{\,}_q = \la^{\,}_h + \la^{\,}_X$;
the Standard Model computation of $\rho(q)$ gives only diagonal
terms [$\rho^{\,}_{++}(q) = \rho^{\,}_{+-;+-}(\qq)$, 
$\rho^{\,}_{--}(q) = \rho^{\,}_{-+;-+}(\qq)$], and one ends up
with the usual probabilistic expression
\beq
\rho^{\,}_{\la^{\,}_h \la^{\,}_h}(h) 
= {1\over N_h} \sum_{q, \la^{\,}_q}
\rho^{\,}_{\la^{\,}_q \la^{\,}_q} \>\>
D_{q,\la^{\,}_q}^{h, \la^{\,}_h} \,,
\label{rhohp2}
\eeq
where $D_{q,\la^{\,}_q}^{h, \la^{\,}_h}$ is the polarized fragmentation 
function of a $q$ with helicity $\la^{\,}_q$ into a hadron $h$ with 
helicity $\la^{\,}_h$. Off-diagonal elements of $\rho(h)$ are all zero.

\goodbreak
\vskip 12pt
\noindent
{\mbox{\boldmath $e^- e^+ \to BX, \> (S_{_B} = 1/2, \> p_{_T}/\sqrt s \to 0)$}}
\vskip 6pt
\nobreak
Let us consider first the case in which $h$ is a spin 1/2 baryon. It was
shown in Ref. \cite{aamr} that in such a case the coherent quark 
fragmentation only induces small corrections to the usual incoherent 
description 
\barr
\rho^{\,}_{++}(B) &=& {1 \over N_{_B}} \sum_q 
\left[ \rho^{\,}_{+-;+-}(\qq) \> D_{q,+}^{B,+} + \rho^{\,}_{-+;-+}(\qq) \> 
D_{q,-}^{B,+} \right] \\
\rho^{\,}_{+-}(B) &=& {\cal O} \left[ \left(
{p_{_T} \over z \sqrt s} \right) \right] \label{rho+-b}\,.
\earr

That is, the diagonal elements of $\rho(B)$ are the same as those given by 
the usual probabilistic formula (\ref{rhohp2}), with small corrections
of the order of $(p_{_T}/z\sqrt s)^2$, while off-diagonal elements are 
of the order $(p_{_T}/z\sqrt s)$ and vanish in the $p_{_T}/\sqrt s \to 0$
limit.

The matrix elements of $\rho(B)$ are related to the longitudinal ($P_z$)
and transverse ($P_y$) polarization of the baryon:
\beq
P_z = 2\rho_{++} - 1, \quad\quad\quad\quad P_y = -2\,{\rm Im} \rho_{+-} \,.
\eeq
Some data are available on $\Lambda$ polarization, both longitudinal and
transverse, from ALEPH Collaboration \cite{aleph} and they do agree with the
above equations. In particular the transverse polarization, at 
$\sqrt s = M_{_Z}$, $p_{_T} \simeq 0.5$ GeV/$c$ and $z \simeq 0.5$ is 
indeed of the order 1\%, as expected from Eq. (\ref{rho+-b}).

\goodbreak
\vskip 12pt
\noindent
{\mbox{\boldmath $e^- e^+ \to VX, \> (S_{_V} = 1, \> p_{_T}/\sqrt s \to 0)$}}
\vskip 6pt
\nobreak
In case of final spin 1 vector mesons one has, always in the limit of small
$p_{_T}$ \cite{akp}, \cite{abmq} 
\barr
\rho^{\,}_{00}(V) &=& {1 \over N_{_V}} \sum_q D_{q,+}^{V,0} \\
\rho^{\,}_{11}(V) &=& {1 \over N_{_V}} \sum_q
\left[ \rho_{+-;+-}(\qq) D_{q,+}^{V,1} + \rho_{-+;-+}(\qq) D_{q,-}^{V,1}
\right] \\
\rho^{\,}_{1,-1}(V) &=& {1 \over N_{_V}} \sum_{q,X}
D^{\,}_{10;+-} \> D^*_{-10;-+} \> \rho^{\,}_{+-;+-}(\qq) \,.
\earr

Again, the diagonal elements have the usual probabilistic expression; 
however, there is now an off-diagonal element, $\rho^{\,}_{1,-1}$, 
which may survive even in the $p_{_T}/\sqrt s \to 0$ limit. In the sequel
we shall concentrate on it. Let us first notice that, in the collinear limit,
one has
\barr
D^{V,0}_{q,+} &=& \sum_X \vert D^{\,}_{0-1;+-} \vert^2 = D^{V,0}_{q,-} \\
D^{V,1}_{q,+} &=& \sum_X \vert D^{\,}_{10;+-} \vert^2 = D^{V,-1}_{q,-} \\
D^{V,1}_{q,-} &=& \sum_X \vert D^{\,}_{12;+-} \vert^2 = D^{V,-1}_{q,+} \,,
\earr
with $ D_q^V = D^{V,0}_{q,+} + D^{V,1}_{q,+} + D^{V,-1}_{q,+}$ and 
$N_{_V} = \sum_q D_q^V$. We also notice that the two fragmentation 
amplitudes appearing in Eq. (30) are related by parity and their product
is always real. $\rho^{\,}_{00}$ and $\rho^{\,}_{1,-1}$ can be measured
through the angular distribution of two body decays of $V$. 

In order to give numerical estimates of $\rho^{\,}_{1,-1}$ we make some
plausible assumptions
\beq
D^{h,1}_{q,-} = D^{h,-1}_{q,+} = 0 \quad\quad\quad
D^{h,0}_{q,+} = \alpha^V_q \> D^{h,1}_{q,+} \label{ass} \,.
\eeq
The first of these assumptions simply means that quarks with helicity
1/2 ($-1/2$) cannot fragment into vector mesons with helicity $-1$ ($+1$).
This is true for valence quarks assuming vector meson wave functions 
with no orbital angular momentum, like in $SU(6)$. The second assumption 
is also true in $SU(6)$ with $\alpha^V_q = 1/2$ for 
any valence $q$ and $V$. Rather than taking  
$\alpha^V_q = 1/2$ we prefer to relate the value of $\alpha^V_q$ to the 
value of $\rho^{\,}_{00}(V)$ which can be or has been measured.
In fact, always in the $p_{_T} \to 0$ limit, one has \cite{abmq}
\beq
\rho^{\,}_{00}(V) = {\sum_q \alpha^V_q \, D^{h,1}_{q,+}
\over \sum_q \> (1+\alpha^V_q) \, D^{h,1}_{q,+}} \,\cdot
\label{rho00}
\eeq
If $\alpha^V_q$ is the same for all valence quarks in $V$ 
($\alpha^V_q = \alpha^V$) 
one has, for the valence quark contribution:
\beq
\alpha^V = {\rho^{\,}_{00}(V) \over 1 - \rho^{\,}_{00}(V)} \,\cdot
\label{alrho}
\eeq

Finally, one obtains \cite{abmq}
\beq
\rho^{\,}_{1,-1}(V) \simeq [1 - \rho^{\,}_{0,0}(V)] \,
{\sum_q \, D^{V,1}_{q,+} \> \rho_{+-;-+}(\qq) 
\over \sum_q \, D^{V,1}_{q,+}} \,\cdot
\label{rho1-1tss}
\eeq

We shall now consider some specific cases in which we expect 
Eq. (\ref{rho1-1tss}) to hold; let us remind once more that our 
conclusions apply to spin 1 vector mesons produced in 
$e^- e^+ \to q \bar q \to V+X$ processes in the limit of small $p_{_T}$
and large $z$, {\it i.e.}, to vector mesons produced in two jet events
($e^- e^+ \to \qq$) and collinear with one of them ($p_{_T} = 0$), 
which is the jet generated by a quark which is a valence quark for the
observed vector meson (large $z$). These conditions should be met 
more easily in the production of heavy vector mesons. 

Among other results one obtains \cite{abmq}:
\barr
\rho^{\,}_{1,-1}(D^{*}) &\simeq& [1 - \rho^{\,}_{0,0}(D^{*})] \>
\rho_{+-;-+}(c \bar c) \label{rhoda} \\
\rho^{\,}_{1,-1}(\phi)  &\simeq& [1 - \rho^{\,}_{0,0}(\phi)] \>
\rho_{+-;-+}(s\bar s) \label{rhopa} \\
\rho^{\,}_{1,-1}(K^{*0}) &\simeq& {1\over 2} \> [1 - \rho^{\,}_{0,0}(K^{*0})] 
\> [\rho_{+-;-+}(d\bar d) + \rho_{+-;-+}(s\bar s)] \,. \label{rhok0a} 
\earr

Eqs. (\ref{rhoda})-(\ref{rhok0a}) show how the value of $\rho^{\,}_{1,-1}(V)$
are simply related to the off-diagonal helicity density matrix element 
$\rho_{+-;-+}(\qq)$ of the $\qq$ pair created in the elementary 
$e^- e^+ \to \qq$ process; such off-diagonal elements would not appear 
in the incoherent independent fragmentation of a single quark, yielding 
$\rho^{\,}_{1,-1}(V)=0$.

By inserting into the above equations the value of $\rho^{\,}_{00}$ when
available \cite{opal} and the expressions of $\rho^{\,}_{+-;-+}$,
Eq. (18), one has:
\barr
\rho^{\,}_{1,-1}(D^{*}) &\simeq& -(0.216 \pm 0.007) \ 
{\sin^2\theta \over 1 + \cos^2\theta} \\
\rho^{\,}_{1,-1}(\phi) &\simeq& -(0.078 \pm 0.014) \ 
{\sin^2\theta \over 1 + \cos^2\theta} \\
\rho^{\,}_{1,-1}(K^{*0}) &\simeq& -0.170 \ 
[1- \rho^{\,}_{0,0}(K^{*0})] \ {\sin^2\theta \over 1 + \cos^2\theta}  \,\cdot
\earr
Finally, in case one collects all meson produced at different angles in
the full available $\theta$ range (say $\alpha < \theta < \pi -\alpha, 
\> |\cos\theta| < \cos\alpha$) an average should be taken in $\theta$, 
weighting the different values of $\rho^{\,}_{1,-1}(\theta)$ with the 
cross-section for the $e^-e^+ \to V+X$ process; this gives \cite{abmq}:
\barr
\langle \rho^{\,}_{1,-1}(D^{*}) \rangle_{[\alpha, \pi-\alpha]}
&\simeq& -(0.216 \pm 0.007) \ 
{3 - \cos^2\alpha \over 3 + \cos^2\alpha} \label{rhodn} \\
\langle \rho^{\,}_{1,-1}(\phi) \rangle_{[\alpha, \pi-\alpha]}
&\simeq& -(0.078 \pm 0.014) \ 
{3 - \cos^2\alpha \over 3 + \cos^2\alpha} \label{rhopn} \\
\langle \rho^{\,}_{1,-1}(K^{*0}) \rangle_{[\alpha, \pi-\alpha]}
&\simeq& -0.170 \ [1- \rho^{\,}_{0,0}(K^*)] \ 
{3 - \cos^2\alpha \over 3 + \cos^2\alpha} \,\cdot \label{rhok0n} 
\earr

These results have to be compared with data \cite{opal}
\barr
\rho^{\,}_{1,-1}(D^*) &=& -0.039 \pm 0.016 \quad\quad {\rm for}
\quad\quad z > 0.5 \quad\quad \cos\alpha = 0.9 \\ 
\rho^{\,}_{1,-1}(\phi) &=& -0.110 \pm 0.070 \quad\quad {\rm for}
\quad\quad z > 0.7 \quad\quad \cos\alpha = 0.9 \\ 
\rho^{\,}_{1,-1}(K^{*0}) &=& -0.090 \pm 0.030 \quad\quad {\rm for}
\quad\quad z > 0.3 \quad\quad \cos\alpha = 0.9  
\earr
which shows a good qualitative agreement with the theoretical
predictions. We notice that while the mere fact that $\rho_{1,-1}$ differs 
from zero is due to a coherent fragmentation of the $\qq$ pair, the actual 
numerical values depend on the Standard Model coupling constants; for example,
$\rho_{1,-1}$ would be positive at smaller energies, at which the one gamma
exchange dominates, while it is negative at LEP energy where the one $Z$
exchange dominates. $\rho_{1,-1}$ has also a peculiar dependence on 
the meson production angle, being small at small and large angles
and maximum at $\theta = \pi/2$. Such angular dependence has been tested
in case of $K^{*0}$ production, assuming no dependence of $\rho_{00}$
on $\theta$, and indeed one has \cite{opal}, in agreement 
with Eqs. (\ref{rhok0a}) and (\ref{rho+-ap})
\beq
\left[ {\rho^{\,}_{1,-1} \over 1- \rho^{\,}_{00}} \right]_{|\cos\theta|<0.5} 
\cdot 
\left[ {\rho^{\,}_{1,-1} \over 1- \rho^{\,}_{00}} \right]^{-1}
_{|\cos\theta|>0.5} = 1.5 \pm 0.7
\eeq

These results are encouraging; it would be interesting to have more and 
more detailed data, possibly with a selection of final hadrons with the 
required features for our results to hold. It would also be interesting to 
test the coherent fragmentation of quarks in other processes, like
$\gamma\gamma \to VX$, $pp \to D^*X$ and $\ell p \to VX$. The first two
processes are similar to $e^-e^+ \to VX$ in that a $\qq$ pair is created 
which then fragments coherently into the observed vector meson; one assumes
that the dominating elementary process in $pp \to D^*X$ is $gg \to c \bar c$.
In both these cases one has for $\rho^{\,}_{+-;-+}(\qq)$ the same value
as in Eq. (\ref{rhoqqelm}), so that one expects a {\it positive} value
of $\rho^{\,}_{1,-1}(V)$. 

In the last process, the production of vector mesons in DIS, the quark 
fragmentation is in general a more complicated interaction of the quark 
with the remnants of the proton and it might be more difficult to obtain 
numerical predictions. However, if one observes $D^*$ mesons one can 
assume or select kinematical regions for which the underlying elementary 
interaction is $\gamma^* g \to c\bar c$: again, one would have the same 
$\rho^{\,}_{+-;-+}(c\bar c)$ as in Eq. (\ref{rhoqqelm}), and one would 
expect a positive value of $\rho^{\,}_{1,-1}(D^*)$. It would indeed be
interesting to perform these simple tests of coherent fragmentation 
effects.

\vskip 18pt
\noindent
{\bf Acknowledgements}
\vskip 6pt
\noindent
I would like to thank the organizers of the Workshop and DESY for financial
support

\end{document}